\documentclass[aps,12pt,eqsecnum,tightenlines,showpacs,nofootinbib,endfloats,amsmath,amssymb]{revtex4}
\usepackage{graphicx}
\usepackage{bm}

\begin{document}

\topmargin -2cm
\author{A.A. Garcia--Diaz}
 \altaffiliation{aagarcia@fis.cinvestav.mx}
\author {G. Gutierrez--Cano }
 \affiliation{Departments~de~F\'{\i}sica,
 ~Centro~de~Investigaci\'on~y~de~Estudios~Avanzados~del~IPN,\\
 Apdu. Postal 14-740, 07000 M\'exico DF, M\'exico.\\}

 \title{Dilaton minimally coupled to $2+1$ Einstein--Maxwell fields;
 stationary cyclic symmetric black holes}
 \date{\today}

\begin{abstract}

Using the Schwarzschild coordinate frame for a static cyclic
symmetric metric in $2+1$ Einstein gravity coupled to a electric
Maxwell field and a dilaton logarithmically depending on the radial
coordinate in the presence of an exponential potential the general
solution of the Einstein--Maxwell--dilaton equations is derived and
it is identified with the Chan--Mann charged dilaton solution. Via a
general $SL(2,R)$--transformation, applied on the obtained charged
dilaton metric, a family of stationary dilaton solutions has been
generated; these solutions possess five  parameters: dilaton and
cosmological constants , charge, momentum, and mass for some values
of them.
 All the exhibited solutions have been characterized
by their quasi-local energy, mass, and momentum  through their
series expansions at spatial infinity. The structural functions
determining these solutions increase as the radial coordinate does,
hence they do not exhibit an dS--AdS behavior at infinity
  Moreover, the algebraic structure of the
Maxwell field, energy-momentum, and Cotton tensors is given explicitly.

\vspace{0.5cm}
\end{abstract}

\maketitle\tableofcontents

 \section{Introduction}

 During the last two decades three--dimensional gravity has received
 some attention, in particular, in topics such as: black hole
 physics, search of exact solutions, quantization of fields coupled
 to gravity, cosmology, topological aspects, and others. This
 interest in part has been motivated by the discovery, in 1992, of
 the $2+1$ stationary circularly symmetric Ba\~nados--Teileboim--Zanelli
 (BTZ) anti-de Sitter black hole ~\cite{BTZ},
 see also \cite{Banados:1992gq,Cangemi:1992my,Carlip:1995qv}, which possesses
 certain features inherent to $3+1$ black holes. On the other hand,
 it is believed that $2+1$ gravity may provide new insights towards a
 better understanding of the physics of $3+1$ quantum gravity.
 The list of references on the mentioned--above topics is rather vast.
 In particular, on exact solutions in $2+1$ gravity
 one finds works on point masses,
 cosmological universes, perfect fluid
 solutions, dilaton fields, scalar fields minimally
 and non--minimally coupled to gravity,
 electromagnetic fields, among others.
 Nevertheless, the literature on stationary
 rotating scalar field solutions
 is rather scarce~\cite{Chan:1994qa,Chan:1995wj},
  most of the known solutions
 of this class are static~\cite{Chan:1996rd,Martinez:1999qi,AyonBeato:2001sb}.

 The purpose of this contribution is to determine families
 of static and stationary cyclic
 symmetric solutions of the $2+1$ gravity, in the
 Schwarzschild coordinate frame, for a minimally
 coupled dilaton field depending logarithmically
 on radial coordinate and
 allowing for an exponential potential in
 the presence of an electromagnetic Maxwell field.
 The static cyclic
 symmetric metric can always be described by
 three structural functions
 $ds^2=-A(r)^2dt^2+B(r)^2dr^2+C(r)^2dt^2$,
 leaving still a freedom in the choice of the $r$--coordinate,
 which can be used to fix the metric structure, for instance:\\
 the Schwarzschild coordinate
 frame: $ds^2=-A(r)^2dt^2+B(r)^2dr^2+r^2dt^2,$\, or
 the $g_{tt}=-1/g_{rr}$ coordinate frame:
 $ds^2=-F(r)^2dt^2+dr^2/F(r)^2+H(r)^2dt^2.$
 In one of the well-known works on
 dilaton~\cite{Chan:1994qa}, this last metric was used
and it was also assumed $H(r)=r^{N/2}$ making troublesome the
establishing of uniqueness of the derived solutions under certain
conditions.

The Schwarzschild coordinate frame results adequate when searching
for electric solutions within Maxwell electrodynamics, nevertheless
it yields to troubles when dealing with magnetic fields. On these
lines of thinking, integrating for Maxwell fields,
Peldan~\cite{Peldan:1992mp}
 succeeded to integrate for the first
 time in the usual Schwarzschild gauge
 the electrostatic Einstein--Maxwell
 field equations with cosmological constant.
 Nevertheless, in
 the search for a
 magnetic solution, Peldan's Section 5.1.2,
 one finds the following comment: ``the reason why I
 did not choose the Schwarzschild gauge from
 the beginning, is that in choosing
 that gauge, I not have been able to
 find an explicit solution for the metric".
 The integration was accomplished in
 the $g_{tt}=-1/g_{rr}$--gauge, leaving the remaining
 metric function free. \\

This paper is organized as follows. Section \ref{ScalarStatic} is
devoted to the derivation of the general static cyclic symmetric
solution for a charged dilaton field in the Schwarzschild coordinate
 frame, $g_{\theta\theta}=r^2$; this
 solution is equipped with four parameters:
cosmological and dilaton constants, mass, and charge. It should be
pointed out that this solution occurs to be equivalent to the
charged static dilaton Chan-Mann solution~\cite{Chan:1995wj}, see
 Sec.~\ref{ScalarROTChan}. The evaluation of the quasi
 local mass is carried out.
 The classification of the Maxwell field and
 energy momentum tensor is accomplished, as well as the
 algebraic characterization of the Cotton tensor is done.

 In Section \ref{ScalarROT}, using $SL(2,R)$--
 transformations of the Killing coordinates, a
 family of rotating metrics coupled to a dilaton field
 is generated. A particular transformation
 gives rise to a charged generalization of the rotating
 Chan--Mann solution~\cite{Chan:1995wj}, see
 Sec.~\ref{ScalarROTChan}.
 For this class solutions the determination of
 the quasi local energy, mass, and momentum
 is carried out. This family of solution is
 endowed with five parameters; in particular
 the allow the interpretation of mass,
 angular momentum,charge, dilaton and
 cosmological constant. Moreover,
 the algebraic classification of the field, energy--momentum and
 Cotton tensors is accomplished.
 A summary of results is presented
 in Concluding Remarks~\ref{Remarks}.

 \subsection{\label{sec:cyclic} Einstein--Maxwell--scalar
 field equations }
 The action to be considered in this work
 dealing (2+1)-dimensional gravity is given by
 \begin{equation}\label{action}
 \mathcal{S}=\int{d^3x}\sqrt{-g}\left[
 R-\frac{B}{2}\nabla_{\mu}{\Psi}\,
 \nabla^{\mu}{\Psi}+2\,{\rm e}^{b\Psi}\Lambda
 -{\rm e}^{-4\,a\,\Psi}\,{ F^2}\right],
 \end{equation}
 where $\Lambda, b$ are arbitrary at this stage
 parameters, ${\Psi}$ is the massless
 minimally coupled scalar field, $R$ is the
 scalar curvature, and  $F^2=F_{\mu\,\nu}\,F^{\mu\,\nu}$
 the electromagnetic invariant.
 The variations of this action yield the dynamical equations
  \begin{eqnarray}\label{Einsteinfield}
 &&{R_{\mu\nu}}= \frac{B}{2}
 \nabla_{\mu}{\Psi}\,\nabla_{\nu}{\Psi}-2{g_{\mu\nu}}{\rm e}^{\,b\,\Psi}\Lambda
 +2\,{\rm e}^{-4\,a\,\Psi}\left({F_\mu}^\alpha\,{F_\nu}_\alpha-g_{\mu\,\nu}\,F^2\right),
 \nonumber\\
 &&
 \frac{B}{2}{\nabla}^{\mu}{\nabla}_{\mu}{{\Psi}}
 +b\,{\rm e}^{\,b\,\Psi}\Lambda+2\,a\,{\rm e}^{-4\,a\,\Psi}\,{ F^2}
 =0, \nonumber\\
 &&
 \nabla^{\mu}\left({\rm e}^{-4\,a\,\Psi}\,F_{\mu\,\nu}\right)=0.
 \end{eqnarray}
 \marginpar{EQ MODI}

 \section{ General static cyclic symmetric black hole solution
 coupled to a scalar field $\Psi(r)=k\,{\rm \ln}(r)$}\label{ScalarStatic}

 The static cyclic symmetric metric in the $2+1$ Schwarzschild coordinate frame is given by
 \begin{equation}\label{metricscLNK_rW_0}
 {g}=-N (r)^2{dt}^2+\frac{{dr}^2}{L(r)^2}
 +r^2{d\phi}^2.
 \end{equation}
 The electromagnetic field equations for the tensor
 field $F_{\mu\nu}=2F_{tr}\delta^t_{[\mu}\delta^r_{\nu]}$,
 and the dilaton $\Phi(r)=k\,{\rm \ln}(r)$ becomes
 \begin{eqnarray}\label{scalarfeqr}
 EQ_F=\frac{d}{dr}\frac {F_{tr} \, L \, r^{-4\,a\,k+1}}{N}
  \rightarrow F_{tr} = {\it Q}\frac{N}{L}{r}^{4\,ak-1}.
 \end{eqnarray}
 The simplest Einstein equations occurs to
 be ${\it R_{11}}+{\it R_{22}}\,L^2\,N^2$, which yields
 \begin{eqnarray}
{\frac {1}{N }}{\frac {d}{dr}}N-{
\frac {1}{L }}{\frac {d}{dr}}L -\frac{1}{2}
\,{\frac {B{k}^{2}}{{r}}}=0,
 \end{eqnarray}
 thus one gets
 \begin{eqnarray}
 N(r)={\it C_N}\,L \left( r \right) {r}^{B\,{k}^{2}/2}.
 \end{eqnarray}
 On the other hand, the equation ${\it R_{33}}$
 gives a first order equation for $L^2=Y \left( r \right)$, namely
 \begin{eqnarray}\label{eqL2sol}
 &&
 {\frac {d}{dr}}Y \left( r \right) +\frac{1}{2}\,{\frac {B{k}^{2}Y \left( r
 \right) }{r}}+2\,{\frac { {r}^{4\,ak} {{\it Q}}^{2}}
 {r}}-2\,\Lambda\,{r}^{bk+1}=0
 \end{eqnarray}
 integrating one obtains
 \begin{eqnarray}\label{L2sol}
 L(r)^2=Y \left( r \right) =-4\,{\frac {{r}^{4\,ak}{{\it Q}}^{2}}{B{k}^{
2}+8\,ak}}+4\,{\frac {\Lambda\,{r}^{2+bk}}{4+\,B{k}^{2}+2bk}}+{r}^{-
1/2\,B{k}^{2}}{\it C_1}
\end{eqnarray}

 Substituting this expression of $Y \left( r \right)$ into the
 remaining scalar field equation
\begin{eqnarray}
 {\frac {d}{dr}}Y \left( r \right) +\frac{1}{2}\,{\frac {B{k}^{2}Y \left( r
 \right) }{r}}-8\,{\frac {
{r}^{4\,ak}\,a{{\it Q}}^{2}}{B\,k\,r}}+2\,{\frac {b\Lambda\,{r}^{bk+1}}{Bk}}=0,
 \end{eqnarray}
 one arrives at
 relationships between constants, namely
 \begin{eqnarray}
 a=-\frac{1}{4\,B\,k},\,b=-B\,k.
 \end{eqnarray}
 \noindent\\
 Therefore, the general charged dilaton  static solution can be given as
 \begin{eqnarray}\label{staticGADx}
 &&
 {g}=-{\it C_N}^{2}\,{r}^{B\,{k}^{2}}
 \,L(r)^2{dt}^2+\frac{{dr}^2}{L(r)^2}+r^2{d\phi}^2,
 \nonumber\\
 &&L(r)^2=\left({r}^{B\,{k}^{2}/2}{\it C_1}
 +4\,{\frac {{r}^{2}\Lambda}{4-B\,{k}^{2}}}+4\,{\frac {{{\it Q}}^{2}}{B{
k}^{2}}}
 \right) {r}^{-B\,{k}^{2}},\,{k}^{2}\neq\frac{4}{B},
 \nonumber\\
 &&
 F_{\mu\nu}=2F_{tr}\delta^t_{[\mu}\delta^r_{\nu]},
 \,F_{tr}={\it Q}{\it C_N}\,{r}^{-1/2\,B{k}^{2}-1}=-A_{t,r}
 \rightarrow\,A_{t}=2\frac{{\it Q}{\it C_N}}{B\,k^2} \,{r}^{-B\,{k}^{2}/2},\nonumber\\
 &&
 \Psi(r)=k\,{\rm \ln}{(r)},
 \end{eqnarray}
 endowed with four relevant parameters: in particular,
 one may identify the mass $M=-{\it C_1}$,
 cosmological constant $\Lambda\rightarrow\pm\frac{1}{l^2}$,
 dilaton parameter  $k$, and the charge $
 {\it Q}$. The constant ${\it C_N}$ can be
 absorbed by scaling
 the coordinate $t$, thus it can be equated
 to unit, ${\it C_N}\rightarrow 1$. Moreover, one has to set
 the charge $Q$ to zero, $Q=0$, when looking
 for the limiting solutions for vanishing dilaton $k=0$, which are
 just the dS and AdS solutions with
 parameters ${\it C_1}=\pm M$ respectively,
 and ${\it C_N}=1$. There is no static electrically
 charged limit of this solution for vanishing dilaton field.

 The constant $\Lambda $ can be
 equated to minus the standard cosmological
 constant $\Lambda_{s}=\pm \frac{1}{l^2}$;
 indeed, by setting in (\ref{staticGADx})

 \begin{eqnarray}
 &&\Lambda=\pm \frac{1}{l^2}\alpha^2,\,r
 \rightarrow r\alpha^{2/(B\,k^2)},\,\phi\rightarrow
 \phi\,\alpha^{-2/(B\,k^2)},{\it Q}\rightarrow
 {\it Q}\,\alpha^{(1+2/(B\,k^2))},
 \nonumber\\
 &&
 {\it C_1}\rightarrow {\it C_1}\,\alpha^{1+4/(B\,k^2)},
 {\it C_N}\rightarrow {\it C_N}\,\alpha^{-(1+2/(B\,k^2))},
\end{eqnarray}
one arrives at the metric (\ref{staticGADx}) with $\Lambda=\pm\frac{1}{l^2}$.
Notice that the $\Lambda$ used by in
Chan--Mann works, when considered as a cosmological
constant, differs from the standard
cosmological constant $\Lambda_{s}=\pm \frac{1}{l^2}=-\Lambda$,
 where $+$ and $-$ stand correspondingly for de Sitter and
 Anti de Sitter (AdS).\noindent\\

If $B>0$, $\Lambda\,\,(4-B\,{k}^{2})\,>0,$
then one has
 \begin{itemize}
 \item a) dS horizonless: $\Lambda<0 \wedge \{k<-\frac{\sqrt{B}}{\sqrt{2}},
 k>\frac{\sqrt{B}}{\sqrt{2}}\}$,${\it C_1}>0,$
 \item b) dS cosmological singularity: $\Lambda<0 \wedge \{k<-\frac{\sqrt{B}}{\sqrt{2}},
 k>\frac{\sqrt{B}}{\sqrt{2}}\}$,${\it C_1}<0,$
  \item c) AdS horizonless: $\Lambda>0 \wedge \{-\frac{\sqrt{B}}{\sqrt{2}}<k
 <\frac{\sqrt{B}}{\sqrt{2}}\}$,${\it C_1}>0,$
 \item d) AdS black hole: $\Lambda>0 \wedge \{-\frac{\sqrt{B}}{\sqrt{2}}<
 k<\frac{\sqrt{B}}{\sqrt{2}}\}$,${\it C_1}<0,$
 \end{itemize}
 If $B>0$, $\Lambda\,\,(4-B\,{k}^{2})\,>0,$ then one has
 \begin{itemize}
 \item e) dS cosmological singularity:
 $\Lambda<0 \wedge \{-\frac{\sqrt{B}}{\sqrt{2}}<k<
 \frac{\sqrt{B}}{\sqrt{2}}\}$,${\it C_1}>0,$
 \item f) AdS event horizon: $\Lambda>0 \wedge \{k<-\frac{\sqrt{B}}{\sqrt{2}},
 k>\frac{\sqrt{B}}{\sqrt{2}}\}$,${\it C_1}>0.$
 \end{itemize}

  The norm of the normal vector
  $n_{\mu}=\nabla_{\mu} r$ to the surface
  $r= r_0=\rm const.$ is given by $g^{rr}(r_0)=L(r_0)^2.$
  It becomes a null vector at the horizon $r_h$ of the
  black hole solution with $B>0$, $k^2<4/B$, ${\it C_1}<0$, and $\Lambda=\frac{1}{l^2}
 $ which is determined as the outer root of $g^{rr}(r)=L(r)^2=0,$ namely
 \begin{eqnarray}
 {r}^{B\,{k}^{2}/2}{\it C_1}
 +4\,{\frac {{r}^{2}\Lambda}{4-B\,{k}^{2}}}+4\,{\frac {{{\it Q}}^{2}}{B{
k}^{2}}}
 =0.
 \end{eqnarray}
 The electromagnetic invariant of this solution amounts to
 $$F_{\mu\nu}F^{\mu\nu}=-2\,{\frac {{{\it Q}}^{2}
 {{\it C_1}}^{2}}{{{\it C_N}}^{2}}}{r}^{-2(1+\,B
{k}^{2})}.
 $$
Accomplishing in the general above
solution (\ref{staticGADx}) the $r$--transformation
and constant parameterizations:
$$r\rightarrow{r}^{N/2},\,k=\pm\frac{\sqrt{2}}{\sqrt{B}}\,\sqrt{\frac{2-N}{N}},$$
one gets, modulo minor constants arrangements,
the Chan--Mann static
solution~\cite{Chan:1994qa,Chan:1995wj}:
\begin{eqnarray}\label{ChanMannSOL}
&&ds^2=-{\it C_N}^2\,U(r){dt}^2+\frac{N^2}{4\,{U(r)}}
{{dr}^2}+r^N{d\phi}^2, \nonumber\\&& U(r)={\it C_1}r^{1-{N}/{2}}
+\frac{2N\Lambda}{(3\,N-2)}r^N+4\frac{N}{2-N}{\it Q}^2,
\nonumber\\&&
\nonumber\\
 &&
 F_{\mu\nu}=2F_{tr}\delta^t_{[\mu}\delta^r_{\nu]},
 \,F_{tr}=\frac{N}{2}{\it Q}{\it C_N}\,{r}^{N/2-2}=-A_{t,r}
 \rightarrow\,A_{t}=\frac{N}{2-N}\frac{{\it Q}{\it C_N}}\,{r}^{N/2-1},\nonumber\\
 &&
\Psi(r)=\pm\sqrt{\frac{N(2-N)}{2\,B}} \,{\rm \ln}{(r)}, \,0<N<2,\,N\neq2/3.
\end{eqnarray}

 \subsection{ Quasi local momentum, energy and mass for the static charged black hole
 solution coupled to a scalar field}\label{ScalarMassStatic}

 To characterize non asymptotically flat solutions one uses
 the Brown--York formalism~\cite{BrownY93}  of quasi--local momentum, energy, and mass
 quantities, which for the stationary cyclic symmetric metric
 \begin{equation}\label{metricscLNKW}
 {g}=-N (r)^2{dt}^2+\frac{{dr}^2}{L(r)^2}
 +K(r)^2({d\phi}+W(r){dt})^2.
 \end{equation}
 are given by:
 \begin{eqnarray}\label{globalJ}
 j_{\phi}=\frac{1}{2\pi}\frac{L}{N}K^2\,W_{,\,r}|_{r\rightarrow
 {\infty}},\,\,J({\partial}/{\partial \phi})=
 \frac{L}{N}K^3\,W_{,\,r}|_{r\rightarrow
 {\infty}},\,J={2\pi}{K}j_{\phi},
 \end{eqnarray}
 \begin{eqnarray}\label{energsurf}
 \epsilon({\partial}/{\partial
 t})=-\frac{1}{\pi\,K}LK_{,\,r}|_{r\rightarrow{\infty}}-\epsilon_{0},
 E({\partial}/{\partial
 t})=-2LK_{,\,r}|_{r\rightarrow{\infty}}
 -2\pi\,\epsilon_{0}\,K|_{r\rightarrow
 {\infty}},
 \end{eqnarray}
 \begin{eqnarray}\label{globalM1}
 M({\partial}/{\partial t})=-2\,NLK_{,\,r}|_{r\rightarrow{\infty}}
 -\frac{L}{N}K^3WW_{,\,r}|_{r\rightarrow{\infty}}-2\pi\,N
 K|_{r\rightarrow{\infty}}\epsilon_{0},
 \end{eqnarray}
 where $\epsilon_{0}$ stands for the energy density of
 the basis metric at infinity, which commonly is the AdS metric with $M$ parameter.
 Incidentally, other useful representation of the mass is
 \begin{eqnarray}\label{globalM}
 M(r)=N(r)E(r)-W(r)J(r).
 \end{eqnarray}
 The evaluation of the mass for the studied charged dilaton solution yields
 \begin{eqnarray}
&& M(r,\epsilon_0) := -2{\it C_N
}\,{r}^{1/2\,B{k}^{2}} L^{2}-2\pi\,\,{r}^{1+1/2\,B{k}^{2}} L\,\epsilon_{0},
\nonumber\\
&&
 M(r,\epsilon_0=0)=-2\,{
\it C_N}\,{\it C_1}-8\,{\frac {{r}^{(4-\,B{k}^{2})/2}{\it C_N}\,\Lambda}{4-B{k}^{2}}}-8\,{
\frac {{r}^{-1/2\,B{k}^{2}}{\it C_N}\,{{\it Q}}^{2}}{B{k}^{2}}}
\end{eqnarray}
Comparing with the energy characteristics of the BTZ
solution, see Appendix~\ref{BTZenergy}, one concludes that role
of the mass parameter is played by ${\it C_1}{\it C_N}=-M$. Recall that ${\it C_N}$
can be equated to $1$. At spatial infinity
all physical quasi--local quantities occur to be infinite, in particular, the field $\Psi$
logarithmically increases.

\subsection{ Algebraic classification of the field, energy--momentum, and
Cotton tensors}\label{RiccCottstatic}

The Maxwell field tensor is given by
\begin{eqnarray}
({F^{\mu}}_{\nu})=   \left[ \begin {array}{ccc} 0&-{\frac
{{\it Q}}{{\it C_N}\, L^{2}}}{r}^{-1-1/2\,B{k}^{2}}&0\\
\noalign{\medskip}-{r}^{-1+1/2\,B{k}^{2}} L^{2}{\it Q}\,{\it C_N}&0&0
\\\noalign{\medskip}0&0&0\end {array} \right] ,
\end{eqnarray}
and is algebraically characterized by the following eigenvectors:
\begin{eqnarray}
&&\lambda_{1}={r}^{-1-B{k}^{2}}{\it Q}
:{\bf V1,2}=[{V^1},-{r}^{1/2\,B{k}^{2}} L^{2}
{\it C_N}\,{ V^1},0],
V_{\mu}V^{\mu}=0,{\bf V1}={\bf N1},\nonumber\\
&&
\lambda_{2}=-{r}^{-1-B{k}^{2}}{\it Q}
:{\bf V1}=[{ V^1},{r}^{1/2\,B{k}^{2}} L^{2}{\it C_N}\,{ V^1}
,0],
V_{\mu}V^{\mu}=0,
,{\bf V2}={\bf N2},
\nonumber\\
&&\lambda_{3}=0:{\bf V3}=[0,0,V^3],
V_{\mu}V^{\mu}={{ V^3}}^{2}{r}^{2},{\bf V3}={\bf S3},
\end{eqnarray}
therefore, its type is $\{N,N,S\}.$
For this class of charged dilatons, the energy--momentum tensor is
described by
\begin{eqnarray}
{T^{\mu}}_{\nu}=2{{\it Q}}^{2}{r}^{-2(1+\,B{k}^{2})}{\delta^{\mu}}_{\phi}{\delta^{\phi}}_{\nu},
\end{eqnarray}
and is algebraically characterized by the following eigenvectors:
\begin{eqnarray}
&&\lambda_{1,2}=0
:{\bf V1,2}=[V1,V2, 0]
,
\mathcal{N}:=V_{\mu}V^{\mu}=-{{ V^1}}^{2}{\it C_N}^{2}\,{r}^{B\,{k}^{2}}
 \,L(r)^2+{\frac {{{
V^2}}^{2}}{L^{2}}}
\nonumber\\
&&\mathcal{}
,{\bf V1,2}={\bf T1,2},{\bf N1,2},{\bf S1,2},\nonumber\\
&&\lambda_{3}=2\,{{\it Q}}^{2}{r}^{-2-2\,B{k}^{2}}
:{\bf V3}=[0,0,{ V^3}],V_{\mu}V^{\mu}={{ V^3}}^{2}{r}^{2},{\bf V3}={\bf S3}.
\end{eqnarray}
Hence depending on the sign of the norm  $\mathcal{N}$ one will have
spacelike, $\mathcal{N}>0$, null, $\mathcal{N}=0$, or timelike, $\mathcal{N}<0$,
eigenvectors: ${\bf V1,2}=\{{\bf S},{\bf N},{\bf T}\}$.
Therefore, in the case of a double root one may choose different vector
components determining, for
instance, one spacelike vector $S$ and the other timelike $T$ or
null $N$ one. Therefore, for the Maxwell energy tensor
one may have the algebraic  types:
$\{2S,S\},\{2N,S\},\{2T,S\}$ and
$\{(S,T),S\},\{(T,T),S\},...,\{(T,N),S\}.$

The Cotton tensor amounts to
\begin{eqnarray}
&&{C^{\mu}}_{\nu}={C^{1}}_{3}{\delta^{\mu}}_{t}{\delta^{\phi}}_{\nu}
+{C^{3}}_{1}{\delta^{\mu}}_{\phi}{\delta^{t}}_{\nu}, \nonumber\\&&
{C^{3}}_{1}=-{\frac {{\it C_N}}{32\,{r}^{4} B{k}^{2}}}\left(
{r}^{-B{k}^{2}/2}{{\it C_1}}^{2}{B}^{2}{k}^{4} \left( 4-B{k}^{2}
 \right) +12\,{r}^{-B{k}^{2}}{{\it Q}}^{2}{\it C_1}\,B{k}^{2}
 \left( 4+B{k}^{2}\right)\right.\nonumber\\&&\left.
+4\,{B}^{2}{r}^ {2-B{k}^{2}}\Lambda\,{\it C_1}\,{k}^{4}  +
64\,{r}^{-3\,B{k}^{2}/2}{{\it Q}}^{4} \left(2+ B{k}^{2}\right)
+64\,B{r}^{2-3\,B{k}^{2}/2}\Lambda\,{{ \it Q}}^{2}{k}^{2}
\frac{\left(2+ B{k}^{2}\right)} {\left( 4-B{k}^{2}
\right)}\right),\nonumber\\&& {C^{1}}_{3} =\frac{1}{32\,{\it
C_N}\,{r}^{2}}\left(B\,{\it C_1}\,{k}^{2} \left( 4-B{k}^{2}
\right){r}^{-B{k}^{2}} +16\,{{\it Q}}^{2} \left(
2+B{k}^{2}\right)\,{ r}^{-3/2\,B{k}^{2}}\right).
\end{eqnarray}
Depending on the signs of ${\it C_1}$ and $\Lambda=\pm 1/l^2$ the
components of ${C^{1}}_{3}$ and ${C^{3}}_{1}$ may be positive or
negative ones, consequently there product could be positive.
Therefore, the Cotton tensor, for both
${C^{1}}_{3}$ and ${C^{3}}_{1}$ positive or negative,
is characterized algebraically by
\begin{eqnarray}
&&\lambda_{1}=0:
{\bf V1}=[0,{ V^2},0],V_{\mu}V^{\mu}
={\frac {{ V^2}^2}{ L ^{2}}},{\bf V1}={\bf S1},N^2
={\it C_N}^{2}\,{r}^{B\,{k}^{2}}
 \,L^2,
\nonumber\\&&
\lambda_{2}=\sqrt{{C^{1}}_{3}{C^{3}}_{1}}:{\bf V2}
=[V^1,0,V^1\,\left(\frac{{C^{3}}_{1}}{{C^{1}}_{3}}\right)^{1/2}]
,\mathcal{N}=V_{\mu}V^{\mu}=-{ V^1}^{2}(N^2\,{C^{1}}_{3}-r^2{C^{3}}_{1})/{C^{1}}_{3}
,\nonumber\\&&{\bf V2}={\bf T2,N2,S2},
\nonumber\\&&
\lambda_{2}=-\sqrt{{C^{1}}_{3}{C^{3}}_{1}}:{\bf V 3}
=[V^1,0,-V^1\,\left(\frac{{C^{3}}_{1}}{{C^{1}}_{3}}\right)^{1/2}]
,V_{\mu}V^{\mu}=-{ V^1}^{2}(N^2\,{C^{1}}_{3}-r^2{C^{3}}_{1})/{C^{1}}_{3}
,\nonumber\\&&{\bf V3}={\bf T3,N3,S3}.
\end{eqnarray}
Depending on the sign of the norm  $\mathcal{N}$ one has
spacelike, $\mathcal{N}>0$, null, $\mathcal{N}=0$, or timelike, $\mathcal{N}<0$,
eigenvectors, denoted by: ${\bf V1,2}=\{{\bf S},{\bf N},{\bf T}\}$.
 Therefore, when the eigenvalues are real, the Cotton tensor
allows for types:
$\{2S,S\},\{2N,S\},\{2T,S\}$ and
$\{(S,T),S\},\{(T,T),S\},...,\{(T,N),S\}.$

In the spacetime region where ${C^{1}}_{3}$ and ${C^{3}}_{1}$
are of opposite signs one of the eigenvalues
becomes imaginary and the following scheme arises
\begin{eqnarray}
&&\lambda_{1}=0:
{\bf V1}=[0,{V^2},0],V_{\mu}V^{\mu}
={\frac {{V^2}}{ L ^{2}}},{\bf V1}={\bf S1},\nonumber\\&&
\lambda_{2}=\bar \lambda_{3}=\sqrt{{C^{1}}_{3}{C^{3}}_{1}}:
{\bf V 2}={\bf Z}=[{V^1},0,{\frac {\sqrt {{C^{3}}_{1}}}
{\sqrt {{C^{1}}_{3}}}}{ V^1}],{\bf V 3}= \bar{\bf Z},
\end{eqnarray}
thus, the algebraic type of the Cotton tensor is $\{S,Z,\bar Z\}$.

\section{Stationary cyclic symmetric black hole solutions
coupled to a scalar field generated via  $SL(2,R)$--transformations}\label{ScalarROT}

Subjecting the metric (\ref{staticGADx}) to a general $SL(2,R)$--transformation
\begin{eqnarray}\label{GENmetric}
&&t=\alpha\,T+\beta\,\Phi,
\nonumber\\&&
\phi=\gamma\,T+\delta\,\Phi, \,\,\Delta:=\alpha\delta-\beta\gamma.
\nonumber\\&&
\end{eqnarray}
one arrives at the stationary metric equipped with rotation
\begin{eqnarray}
{g}&&=-\left({\alpha}^{2}{\it C_N}^{2}
{r}^{B{k}^{2}}L^{2}-\gamma^2\,r^{2} \right){d\,T}^2
+\frac{{dr}^2}{{L}^2} \nonumber\\&& -2\left(\beta\, \alpha\,{{\it
C_N}}^{2}{r}^{B{k }^{2}}
L^{2}-\delta\,\gamma\,{r}^{2}\right){d\,T}{d\,\Phi}+\left({\delta}^{2}{r}^{2}
-{\beta}^{2}{{\it C_N}}^{2}  {r}^{B{k}^{2}}\,{ L}^
{2}\right){d\Phi}^2,\nonumber\\&&
{L}(r)^2=\left({r}^{B\,{k}^{2}/2}{\it C_1}
 +4\,{\frac {{r}^{2}\Lambda}{4-B\,{k}^{2}}}+4\,{\frac {{{\it Q}}^{2}}{B{
k}^{2}}}
 \right){r}^{-B\,{k}^{2}}.
\end{eqnarray}
Using the standard notation, this charged rotating
dilaton solution is given as
\begin{eqnarray}\label{metricscLNKWsl}
&& {g}=-\textsf{N}(r)^2{d\,T}^2+\frac{{dr}^2}{\textsf{L}(r)^2}
+\textsf{K}(r)^2({d\Phi}+\textsf{W}(r){d\,T})^2,\nonumber\\
&&
\textsf{N}(r)^2=\frac { N^{2}{r}^{2}}
{{\delta}^{2}{r}^{2}-{\beta}^{2} N^{2}}\left( \alpha
\,\delta-\beta\,\gamma \right) ^{2},
\nonumber\\&&
N^2:=C_N^2\left({r}^{B\,{k}^{2}/2}{\it C_1}
 +4\,{\frac {{r}^{2}\Lambda}{4-B\,{k}^{2}}}+4\,{\frac {{{\it Q}}^{2}}{B{
k}^{2}}}
 \right) =
C_N^2\,{r}^{B\,{k}^{2}}\textsf{L}(r)^2,
\nonumber\\&&
\textsf{L}(r)^2=\left({r}^{B\,{k}^{2}/2}{\it C_1}
 +4\,{\frac {{r}^{2}\Lambda}{4-B\,{k}^{2}}}+4\,{\frac {{{\it Q}}^{2}}{B{
k}^{2}}}
 \right){r}^{-B\,{k}^{2}},
\nonumber\\&&
\textsf{K}(r)^2=
{\delta}^{2}{r}^{
2}-{\beta}^{2}N^{2},
\nonumber\\&&
\textsf{W}(r)={\frac {\delta\,\gamma\,{r}^{2}-\beta\,\alpha\,N ^{2}}{
{\delta}^{2}{r}^{2}-{\beta}^{2}N^{2}}},\,\beta\neq0\neq\delta,
\nonumber\\&&
\Psi=k\,{\rm ln}(r),
 \nonumber\\
 &&
 F_{\mu\nu}=2F_{Tr}\delta^T_{[\mu}\delta^r_{\nu]}
 +2F_{\Phi\,r}\delta^\Phi_{[\mu}\delta^r_{\nu]},\nonumber\\
 &&
 F_{Tr}=-A_{T,r}=\alpha\,F_{tr}
 =-\alpha\frac{d}{dr}\left(2\frac{{\it Q}{\it C_N}}{B\,k^2}
 \,{r}^{-B\,{k}^{2}/2}\right),\,\,F_{tr}
 ={\it Q}{\it C_N}\,{r}^{-1/2\,B{k}^{2}-1},\nonumber\\
 &&
 F_{\Phi\,r}=-A_{\Phi,r}=\beta\,F_{tr}
 =-\beta\frac{d}{dr}\left(2\frac{{\it Q}
 {\it C_N}}{B\,k^2}\,{r}^{-B\,{k}^{2}/2}\right).
\end{eqnarray}

$$\,F_{tr}={\it Q}{\it C_N}\,{r}^{-1/2\,B{k}^{2}-1}=-A_{t,r}
 \rightarrow\,A_{t}=2\frac{{\it Q}{\it C_N}}{B\,k^2} \,{r}^{-B\,{k}^{2}/2},$$

 It is worthwhile to notice that the structure of the
 electromagnetic energy tensor amounts
 to
 \begin{eqnarray}
  T_{\mu\nu} =2\,{{\it Q}}^{2}{r}^{-2\,B{k}^{2}}\left[
\begin {array}{ccc} \gamma^2&0&\gamma\,
\delta\\\noalign{\medskip}0&0&0\\\noalign{\medskip}\gamma\,\delta&0&\delta^2
\end {array} \right].
\end{eqnarray}
For the AdS ($\Lambda=1/l^2$)--black hole branch;
the constants appearing in the structural functions will be replaced by
$C_N\rightarrow{\sqrt{{1-2\,\,{k}^{2}}}}=
{B},\,{ k}=\,\sqrt {1-\,{B}^{2}}/\sqrt{2}$, $0<k<1/\sqrt{2},1>B>0$, thus
\begin{eqnarray}
N(r)^2=N^2_B&&={r}^2\left(r^{-2\,{B}^{2}}
{\it C_1}\,{B}^{2}+\Lambda\right),\,
\textsf{L}(r)^2=L^2_B={\frac {{\it
N}^2_{B}\,}{{B}^{2}}}{r}^{-4+4\,{B}^{2}}.
\end{eqnarray}

\subsection{Quasi local mass and momentum}
The evaluation of the quasi-local momentum $J(r)$ yields
\begin{eqnarray}
{ J(r)}=J({r\rightarrow\infty})=\frac{1}{2}\,\beta\,
\delta\,{\it C_N}\,{\it C1}\, \left(4- B{k}^{2}\right) +8
\,{\frac {\delta\,{\it C_N}\,\beta\,{{\it Q}}^{2}}
{B{k}^{2}}}{r}^{-\,B{k}^{2}/2}
,
\end{eqnarray}
hence, for positive $B>0$, the contribution of the
electromagnetic field ${\it Q}\neq 0$ to the momentum,
at spatial infinity, disappears and one has
\begin{eqnarray}
J({r\rightarrow\infty})=\frac{1}{2}\,\beta\,\delta\,{\it C_N}
\,{\it C1}\, \left(4- B{k}^{2}\right)=:J_0.
\end{eqnarray}
The evaluation of the quasi local mass yields
\begin{eqnarray}
M(r)&&=\frac{1}{2}\,{\frac {{\it C_N}\,\Delta}{r^3B{k}^{2}
\left( B{k}^{2}-4 \right) }}\frac{M_L}{D}-\frac{1}{2}\,
{\frac {\beta\,\delta\,{\it C_N}\,}{{B}^{2}{k}^{4}{r}^{2
}}}\frac{M_J}{D},
\nonumber\\&&
{}
\nonumber\\&&
M_{L}:=-4\,B{k}^{2}{\it C_1}\,{r}^{3} \left( B{k}^{2}-4 \right)
 \left( P+\Lambda\,{\beta}^{2}{{\it C_N}}^{2}B{k}^{2} \right)
\nonumber\\&&
+4\,r{{\it
Q}}^{2}{\beta}^{2}{{\it C_N}}^{2}{\it C_1}\, \left( B{k}^{2}-4 \right)
^{2}B{k}^{2}+16\,B{k}^{2}{r}^{5-1/2\,B{k}^{2}}\Lambda\, P
 \nonumber\\&&
 +{r}^{1+1/2\,B{k}^{2}}{{\it C_1}}^{2}{\beta}^{2}{{\it C_N}}^{2}
{B}^{2}{k}^{4} \left( B{k}^{2}-4 \right) ^{2}
-16\,{{\it Q}}^{2}{r}^{3
-1/2\,B{k}^{2}} \left( B{k}^{2}-4 \right)  \,{P},
\nonumber\\&&
{}
\nonumber\\&&
M_{J}=\left(16\,B{k}^{2}{{\it Q}}^{2}{r}^{-B{k}
^{2}/2}
-{B}^{2}{\it C_1}\,{k}^{4} \right)\left( \delta\,
\gamma\, \left( B{k}^{2}-4 \right) +4\,\Lambda\,{{\it C_N}}^{2}\alpha\,
\beta \right) \left( B{k}^{2}-4 \right)  \,{r}^{2}
\nonumber\\&&
-64\beta\,
\alpha\,{r}^{-1/2\,B{k}^{2}}{{\it Q}}^{4}\,{{\it C_N}}^{2}
\left( B{k}^{2}-4 \right)+\beta\,\alpha{B}^{2}{k}^
{4}{{\it C_1}}^{2}\,{{\it C_N}}^{2}{r}^{1/2\,B{k}^{2}}
 \left( B{k}^{2}-4 \right) ^{2}
\nonumber\\&&
 +4\,\beta\,\alpha\,B{k}^{2}{\it C_1}\,{{
\it C_N}}^{2}{{\it Q}}^{2} \left( B{k}^{2}-4 \right)  \left( B{k}^{2}-
8 \right) ,
\nonumber\\&&
D= P-4\,{\frac {{\beta}^{2}{{\it C_N}}^{2}{{\it Q}}^{2} \left( B
{k}^{2}-4 \right) }{B{r}^{2}{k}^{2}}}-{{\beta}^{2}{{\it C_N}}^{
2}{r}^{(\,B{k}^{2}-4)/2}{\it C_1}\, \left( B{k}^{2}-4 \right) },
\end{eqnarray}
where $P={\delta}^{2}B{k}^{2}-4\,{\delta}^{2}+4\,{\beta}^{2}{{\it C_N}}^{2}\Lambda.
$
The order zero in the series expansion  of $M(r)$ is given by
\begin{eqnarray}
M_0=
2\,M\,\Delta\,{\frac { \Lambda\,{\beta}^{2}{{
\it C_N}}^{2}B{k}^{2}+{\delta}^{2}B{k}^{2}-4\,{\delta}^{2}  }{{
\delta}^{2}B{k}^{2}-4\,{\delta}^{2}+4\,{\beta}^{2}{{\it C_N}}^{2}
\Lambda}}
-J_0
{\frac { B{k}^{2}\delta\,
\gamma+4\,\beta\,\alpha\,{{\it C_N}}^{2}\Lambda-4\,\delta\,\gamma
  }{{\delta}^{2}B{k}^{2}-4\,{
\delta}^{2}+4\,{\beta}^{2}{{\it C_N}}^{2}\Lambda}},
\end{eqnarray}
where ${\it C_N}\,{\it C1}$  has been replaced by $-M$.
Comparing with the quasi--local BTZ mass $ M_{BTZ}=2\,M-2\,
{{r}^{2}}/{{l}^{2}}$, see details in Appendix \ref{BTZenergy}, when the
rotation vanishes, $\beta=0$, $\Delta=\alpha\delta\rightarrow 1$,
implies that $M_0\rightarrow-{\it C_N}\,{\it C1}=M$.\\
It should be pointed out that, due to the presence of terms with
positive powers of $r$ in the series expansion of M(r), it increases
as $r\rightarrow\infty$, like the $ M_{BTZ}$ does.

Frequently, in the literature one encounters the $SL(2,R)$ transformation
\begin{eqnarray}
t={\frac {T}{\sqrt {1-{\frac {{\omega}^{2}}{{l}^{2}}}}}}-\omega\,
{\frac {\Phi}{\sqrt {1-{\frac {{\omega}^{2}}{{l}^{2}}}}}},\,
\phi=-
\frac{\omega}{{l}^
{2}}\,{\frac {T}{\sqrt {1-{\frac {{\omega}^{2}}{{l}^{2}}}}}}
+{\frac {\Phi}{\sqrt {1-{\frac {{\omega}^{2}}{{l}^{2}}}}}},
\end{eqnarray}
correspondingly, the structural functions (\ref{metricscLNKWsl})
assume the form
\begin{eqnarray}
&&
\textsf{N}(r)^2={\frac { \left( {l}^{2}-{\omega}^{2} \right) {r}^{2} N ^{2}}{{l}^{2}
\left( {r}^{2}-{\omega}^{2}N ^{2} \right) }},\,\textsf{L}(r)=\textsf{L}(r),\,N=N,
\nonumber\\&&
\textsf{W}(r)={\frac {\omega\, \left( N ^{2}{l}^{
2}-{r}^{2} \right) }{{l}^{2} \left( {r}^{2}-{\omega}^{2}N ^{2} \right) }},
\textsf{K}(r)^2={\frac {{l}^{2} \left( {r}^{2}-{\omega}^{2}
 N ^{2} \right) }{{l}^{2}-{\omega}^{2}}}.
\end{eqnarray}
In this parametrization, the quasi--local momentum becomes
$ J_0=-\frac{1}{2}\,{\frac {\omega\,l^2}{{l}^{2}
-{\omega}^{2}}}{\it C_1}{\it C_N}(4-B\,k^2),$ and
vanishes as soon the rotation $\omega$ becomes zero.

The generated stationary metric (\ref{metricscLNKWsl}) gives rise,
among others, to a charged generalization of Chan--Mann rotating dilaton solution,
see below Sec.~{\ref{ScalarROTChan}.

With this result we are giving an answer to the remark contained in
the Conclusions of \cite{Chan:1995wj}:\,``Although the static
charged black solutions of (1.1) exist \cite{Chan:1994qa}, at
present  we are unable to generalize our spinning solution to
charged cases. This endeavor is complicated by the fact that when
one adds  Maxwell fields to a spinning solution, both electric and
magnetic fields must be present...''

\subsection{ Algebraic classification of the field, energy--momentum, and
Cotton tensors}\label{RiccCottstatic}

The Maxwell electromagnetic field tensor amounts to
\begin{eqnarray}
\left({F^{\mu}}_{\nu}\right)= \left[
\begin {array}{ccc} 0&-{\frac {\delta\,{r}^{-1-3/2\,B{k}^{2}}{\it Q}}{
 \Delta {\it C_N}\, L^{2}}}&0\\\noalign{\medskip}-\alpha\,
 L^{2}{\it Q}\,{r}^{-1-1/2\,B{k}^{2}}{\it C_N}&0&-\beta\,L^{2}{\it Q}
\,{r}^{-1-1/2\,B{k}^{2}}{\it C_N}\\\noalign{\medskip}0&{\frac {\gamma
\,{r}^{-1-3/2\,B{k}^{2}}{\it Q}}{\Delta{\it C_N}\, L^{2}}}&0
\end {array} \right],
\end{eqnarray}
and has the following eigenvectors
\begin{eqnarray}
&&\lambda_{1}=0
:{\bf V1}=[-{\frac {\beta\,{ V^3}}{\alpha}},0,{ V^3}],
 V_{\mu}V^{\mu}={\frac {{{\it V3}}^{2}{r}^{2}{\Delta}^{2}}{{\alpha}^{2}}}
,\nonumber\\
&&
\lambda_{2}={\it Q}\,{r}^{-1-B{k}^{2}}:
 {\bf V2}=
[-{r}^{-\,B{k}^{2}/2}{\frac {\delta\,{ V^2}}{\Delta\,L^{2}{\it C_N}
 }},{ V^2},{r}^{-\,B{k}^{2}/2} {\frac {
\gamma\,{V^2}}{\Delta\,L^{2}{\it C_N} }}],
V_{\mu}V^{\mu}=0,{\bf V2}={\bf N2}
,\nonumber\\
&&\lambda_{3}=-{\it Q}\,{r}^{-1-B{k}^{2}}
:
{\bf V3}=[{r}^{-\,B{k}^{2}/2}{\frac {\delta\,{ V^2}}
{\Delta\,L^{2}{\it C_N} }},{V^2},{r}^{-\,B{k}^{2}/2}{\frac {
\gamma\,{V^2}}{\Delta\,L^{2}{\it C_N}  }}]
,
V_{\mu}V^{\mu}=0
,{\bf V3}={\bf N3}.\nonumber\\
\end{eqnarray}
Correspondingly, its type is $\{S,N,N\}$.\\
The electromagnetic energy--momentum tensor is given by
\begin{eqnarray}
 \left( {T^\mu}_{\nu}\right) =-2\,{\frac {{{\it Q}}^{2}{r}^{-(2+2\,B{k}^{2})}}{
\Delta}}
\left[
\begin {array}{ccc} \beta\,\gamma\,&0&\beta\,\delta\\\noalign{\medskip}
0&0&0\\\noalign{\medskip}-\alpha\,\gamma&0&-\alpha\,\delta
\end {array} \right],
\end{eqnarray}
and is algebraically characterized by:
\begin{eqnarray}
&&\lambda_{1,2}=0
:{\bf V1,2}=[{V^1},{V^2},{
 -V^1\gamma/\delta}],
\mathcal{N}= V_{\mu}V^{\mu}=-{\frac {{{\it C_N}}^{2} L^{2}
{r}^{B{k}^{2}}{\Delta}^{2}}{{\delta}^{2}}}{{ V^1}}^{2}+{\frac {{{
 V^2}}^{2}}{ L^{2}}}
,\nonumber\\
&&
\mathcal{N}>0,{\bf V1,2}={\bf S1,2},\mathcal{N}=0,{\bf V1,2}={\bf N1,2},\mathcal{N}<0,{\bf V1,2}={\bf T1},
\nonumber\\
&&\lambda_{3}=2\,{{{{\it Q}}^{2}{r}^{-(2+2\,B{k}^{2})}}}
:{\bf V3}=[-{\frac {\beta\,{ V^3}}{\alpha}},0,{\it
V^3}]
,\nonumber\\&&
V_{\mu}V^{\mu}={\frac {{({\it V^3}})^{2}{r}^{2} \Delta ^{2}}{{\alpha}^{2}}}
,{\bf V3}={\bf S3},
\end{eqnarray}
In the case of a double root, depending on the sign of the
norm $V_{\mu}V^{\mu}$, one will have
spacelike, $\mathcal{N}>0$, null, $\mathcal{N}=0$,
or timelike, $\mathcal{N}<0$,
eigenvectors: ${\bf V1,2}=\{{\bf S1,2},{\bf N1,2},{\bf T1,2}\}$.
Therefore, one may choose different vector
components determining, for
instance, one spacelike vector $S$ and the other
timelike $T$ or
null $N$ one. Therefore,
one may have the algebraic Ricci types:
$\{2S,S\},\{2N,S\},\{2T,S\}$ and $\{(S,T),S\},\{(T,T),S\},...,\{(T,N),S\},$
where parenthesis is used to stand out the
multiplicity of the root under consideration.

\marginpar{MODIF}

The matrix of the Cotton tensor for the rotating charged solution is given by
\begin{eqnarray}
&&({C^{\mu}}_{\nu})= \left[
\begin {array}{ccc} {\frac{-\alpha\,
\beta\,{C^{3}}_{1}+\gamma\,\delta
\,{C^{1}}_{3}}{\Delta}}&0&{\frac
{-{\beta}^{2}{C^{3}}_{1}+{\delta}^{2}
{C^{1}}_{3}}{\Delta}}
\\\noalign{\medskip}0&0&0\\\noalign{\medskip}
-{\frac {-{\alpha}^{2}{C^{3}}_{1}+{\gamma}^{2}{C^{1}}_{3}}{\Delta}}&0&{
-\frac {-\alpha\,\beta\,{C^{3}}_{1}+\gamma\,
\delta\,{C^{1}}_{3}}{\Delta}}\end {array} \right]
\end{eqnarray}
where
\begin{eqnarray}
&&{C^{1}}_{3}
=\frac{1}{32\,{\it C_N}\,{r}^{2}}\left(B\,
{\it C_1}\,{k}^{2} \left( 4-B{k}^{2} \right)
{r}^{-B{k}^{2}} +16\,{{\it Q}}^{2} \left( 2+B{k}^{2}\right)\,{
r}^{-3/2\,B{k}^{2}}\right),\nonumber\\&&
{C^{3}}_{1}=-{\frac {{\it C_N}}{32\,{r}^{4} B{k}^{2}}}\left(
{r}^{-B{k}^{2}/2}{{\it C_1}}^{2}{B}^{2}{k}^{4} \left( 4-B{k}^{2}
 \right) +12\,{r}^{-B{k}^{2}}{{\it Q}}^{2}{\it C_1}\,B{k}^{2}
 \left( 4+B{k}^{2}\right)\right.\nonumber\\&&\left.
+4\,{B}^{2}{r}^
{2-B{k}^{2}}\Lambda\,{\it C_1}\,{k}^{4}  +
64\,{r}^{-3\,B{k}^{2}/2}{{\it Q}}^{4} \left(2+ B{k}^{2}\right)
+64\,B{r}^{2-3\,B{k}^{2}/2}\Lambda\,{{
\it Q}}^{2}{k}^{2} \frac{\left(2+ B{k}^{2}\right)}
{\left( 4-B{k}^{2} \right)}\right).\nonumber\\
\end{eqnarray}
\marginpar{VERSIGNO}
Because of the presence of ${\it C_1}$ and $\Lambda=\pm 1/l^2$
in ${C^{1}}_{3}$ and ${C^{3}}_{1}$, these tensor
components may be positive or negative
quantities in some spacetime regions.
Therefore, in general, the Cotton tensor
can be characterized algebraically by three real eigenvalues

\begin{eqnarray}
&&\lambda_{1}=0:
{\bf V1}=[0,{ V^2},0],V_{\mu}V^{\mu}
={\frac {{ V^2}^2}{ L ^{2}}},{\bf V1}={\bf S1},
\nonumber\\&&
\lambda_{2}=\sqrt{{C^{1}}_{3}{C^{3}}_{1}}:
{\bf V2}=[V^1,0,V^3],\nonumber\\&&
{V^3}={\frac {{\it V1}\, \left( -\alpha\,\beta\,{C^{3}}_{1}
+\gamma\,
\delta\,{C^{1}}_{3}
-\sqrt {{C^{3}}_{1}\,{C^{1}}_{3}}\Delta\right) }
{{\delta}^{2}{C^{1}}_{3} -{\beta}^{2}{C^{3}}_{1}}},
\nonumber\\&&
V_{\mu}V^{\mu}={\frac {{{\it V1}}^{2} \Delta^{2}
 \left( {C^{3}}_{1}\,{r}^{2}-{C^{1}}_{3}\, N^{2} \right)
 \left( {\delta}^{2}{C^{1}}_{3}+{\beta}^{
2}{C^{3}}_{1}+2\,\delta\,\beta\sqrt {{C^{3}}_{1}\,{C^{1}}_{3}} \right) }{
 \left( {\delta}^{2}{C^{1}}_{3} -{\beta}^{2}{C^{3}}_{1} \right) ^{2}}},
\nonumber\\&&{\bf V2}={\bf T2,N2,S2},
\nonumber\\&&
\lambda_{3}=-\sqrt{{C^{1}}_{3}{C^{3}}_{1}}:{\bf V 3}
=[V^1,0,V^3]\nonumber\\&&
V^3={\frac {{\it V1}\, \left( -\alpha\,\beta\,{C^{3}}_{1}
+\gamma\,
\delta\,{C^{1}}_{3}+\sqrt {{C^{3}}_{1}\,{C^{1}}_{3}}
\Delta\right) }{{\delta}^{2}{C^{1}}_{3} -{\beta}^{2}{C^{3}}_{1}}}
,
\nonumber\\&&
V_{\mu}V^{\mu}={\frac {{{\it V1}}^{2}\Delta^{2}
\left( {C^{3}}_{1}\,{r}^{2}
-{C^{1}}_{3}\, N^{2} \right)  \left( {\delta}^{2}{C^{1}}_{3}+
{\beta}^{2}{C^{3}}_{1}-2\,\delta\,\beta\sqrt {{C^{3}}_{1}\,{C^{1}}_{3}}
 \right) }{ \left({\delta}^{2}{C^{1}}_{3} -{\beta}^{2}{C^{3}}_{1}
 \right) ^{2}}},
\nonumber\\&&{\bf V3}={\bf T3,N3,S3}.
\end{eqnarray}
In this case the algebraic types of the Cotton tensor are given
by $\{T,T,S\}$, $\{T,S,S\}$,..., $\{S,S,S\}.$\\
For one real and two complex conjugate eigenvalues, one has
\begin{eqnarray}
&&\lambda_{1}=0:
{\bf V1}=[0,{ V^2},0],V_{\mu}V^{\mu}
={\frac {{ V^2}}{ L ^{2}}},{\bf V1}={\bf S1},\nonumber\\&&
\lambda_{2}=\bar \lambda_{3}=\sqrt{{C^{1}}_{3}{C^{3}}_{1}}:
\nonumber\\&&
{\bf V 2}={\bf Z}=[{ V^1},0, V^3],\,{V^3}
=-{\frac {{\it V1}\, \left( -\alpha\,\beta\,{C^{3}}_{1}
+\gamma\,
\delta\,{C^{1}}_{3}-\sqrt {{C^{3}}_{1}\,{C^{1}}_{3}}
\Delta\right) }{{\delta}^{2}{C^{1}}_{3}-{\beta}^{2}{C^{3}}_{1}}},
\nonumber\\&&
{\bf V 3}= \bar{\bf Z}.
\end{eqnarray}
Thus, the algebraic type of the Cotton tensor is given by $\{S,Z,\bar Z\}$.
\marginpar{ModifERROTR}

\subsection{ Particular stationary cyclic symmetric black hole solution
coupled to a scalar field via $SL(2,R)$--transformation}\label{ScalarROTChan}

Requiring the fulfillment of the relationship
\marginpar{Check}
\begin{eqnarray}
\gamma\,\delta=-4{\frac {\,{{\it C_N}}^{2}\Lambda\,}
{ \left( B\,\,{k}^{2}-4 \right) }}\alpha\,\beta\,
\end{eqnarray}
the term with power $r^2$ in $W$ or equivalently in $g_{T\Phi}$
disappears. The generalized rotating charged Chan--Mann metric is given by
\begin{eqnarray}
{g}&&=-\left({\it C_N}^{2}{\alpha}^{2} L^{2}{r}^{B{k}^{2}}
-16\,{\beta}^{2}{\frac {{{\it
C_N}}^{4}{\alpha}^{2}{\Lambda}^{2}{r}^{2}
 }{{\delta}^{2} \left( B{k}^{2}-4 \right) ^{2}}}\right){d\,T}^2
+\frac{{dr}^2}{{L}^2} \nonumber\\&& -2\alpha\,\beta{\frac {  {{\it
C_N}}^{2}}{B{k}^{2}}} \left( {r}^{\,B\,{k}^{2}/2}{\it
C_1}\,B{k}^{2}+4\,{{\it Q}} ^{2}
\right){d\,T}{d\,\Phi}+\left({\delta}^{2}{r}^{2} -{\beta}^{2}{{\it
C_N}}^{2}  {r}^{B{k}^{2}}\,{ L}^ {2}\right){d\Phi}^2, \nonumber\\&&
{L}^2=\left({r}^{B\,{k}^{2}/2}{\it C_1}
 +4\,{\frac {{r}^{2}\Lambda}{4-B\,{k}^{2}}}+4\,{\frac {{{\it Q}}^{2}}{B{
k}^{2}}}
 \right){r}^{-B\,{k}^{2}},\,\Psi=k\,{\rm ln}(r),
\nonumber\\
 &&
 F_{\mu\nu}=2F_{Tr}\delta^T_{[\mu}\delta^r_{\nu]}
 +2F_{\Phi\,r}\delta^\Phi_{[\mu}\delta^r_{\nu]},
 F_{Tr}=\alpha\,F_{tr},\,F_{\Phi\,r}=\beta\,F_{tr},
 \,F_{tr}={\it Q}{\it C_N}\,{r}^{-1-B\,{k}^{2}/2}.\nonumber\\
\end{eqnarray}
Switching of the rotation $\beta=0$, the presented above metric
reduces to the Chan--Mann charged dilaton metric.\\

\section{Concluding Remarks}\label{Remarks}
Using the Schwarzschild coordinate frame for a static cyclic
symmetric metric in $2+1$ gravity coupled minimally to a dilaton
logarithmically depending on the radial coordinate in the presence
of a exponential potential and an electrical Maxwell field the
general solution of the Einstein--Maxwell--dilaton equations is
derived avoiding any ansatz. This static solution occurs to be
equivalent to the Chan--Mann charged dilaton static solution. Via a
general $SL(2,R)$--transformation of the Killing coordinates,
applied on the derived charged static cyclic symmetric metric, a
family of stationary dilaton solutions has been generated; they are
equipped with five relevant parameters interpretable as dilaton
parameter, charge, momentum, cosmological constant, and  mass for
some values of them. A particular $SL(2,R)$--transformation is
identified, which gives raise to the charged generalization of the
rotating Chan--Mann dilaton solution. At spatial infinity all these
solutions do not allow for an AdS-dS limit, there structural
functions increase indefinitely as the radial coordinate increases.
There exists a horizon, structurally common to the full class of
solutions, determining their black hole character for a range of the
physical parameters. This families of solutions, the static and
stationary ones, have been characterized by their quasi-local
energy, mass, and momentum through their series expansions at
spatial infinity. The algebraic classifications of the
electromagnetic field, Maxwell energy--momentum, and Cotton tensors
are established. The electromagnetic field tensor belongs to the
type $\{S,N,N\}$ The Maxwell energy--momentum tensor is of
types:$\{2S,S\},\{2N,S\},\{2T,S\},\{(S,T),S\}
,\{(T,T),S\},...,\{(T,N),S\}.$ The Cotton tensor exhibits various
possibilities. For real roots it falls into types:
$\{T,T,S\},\{T,S,S\},\{T,N,S\},\{S,T,S\}, \{S,N,S\},...,\{S,S,S\}.$
For one complex root, the Cotton tensor is of the type $\{S,Z,\bar
Z\}$.

\appendix

\section{Momentum, energy and mass of the Ba\~nados--Teitelboim--Zanelli black hole}\label{BTZenergy}

Let us consider the asymptotically
anti--de Sitter $(2+1)$--dimensional stationary black hole solution--the BTZ--metric--
given by
\begin{eqnarray}\label{BTZmetric}
&&ds^2=-N(\rho)^2dt^2+\frac{1}{L(\rho)^{2}}\,d\rho^2
+\rho^2[d\phi+W(\rho)dt]^2,\nonumber\\
&&N^2(\rho)=L^{2}(\rho)=-M+\frac{\rho^2}{l^2}
+\frac{J^2}{4\,\rho^2},\,K(\rho)=\rho,
\,W(\rho)=-\frac{J}{2\,\rho^2}.
\end{eqnarray}
The corresponding surface energy and momentum densities, at $\rho=R=\rm costant$, are equal to
\begin{eqnarray}\label{BTZj}
&&\epsilon(R,\epsilon_{0})=-\frac{1}{\pi\,R}\sqrt{-M+\frac{R^2}{l^2}+\frac{J^2}{4\,R^2}}-\epsilon_{0},
\,j_{\phi}(R)=\frac{1}{2\pi}\frac{J}{R}.
\end{eqnarray}
Consequently the total momentum, energy, and mass are
\begin{eqnarray}\label{BTZenergyJ}
&&
J({\partial}/{\partial \phi})=J,
\nonumber\\&&
E(R,\epsilon_{0})=-2\sqrt{-M+\frac{R^2}{l^2}
+\frac{J^2}{4\,R^2}}-2\,\pi\,\epsilon_{0},
\nonumber\\&&
M({\partial}/{\partial
t})=N(R)\,E(R,\epsilon_{0})+\frac{J^2}{2\,R^2}=2M-2\frac{R^2}{l^2}-2\,\pi
\,\epsilon_{0}\sqrt{-M+\frac{R^2}{l^2}+\frac{J^2}{4\,R^2}}.
\end{eqnarray}
These  expressions for surface densities and global quantities are
in full agreement with the corresponding ones reported in
Ref.~\cite{Brown:1994gs}, section IV.\\
Notice that the series expansion of the energy and mass independent
of $\epsilon_{0}$ behave at infinity $R$, which will be denoted from
now on by the same coordinate Greek letter $\rho$ accompanied by
$\rightarrow\infty$ and the approximation sign $\approx$, as
\begin{eqnarray}\label{enerBTZapproxEpsBTZ}
\epsilon(\rho\rightarrow\infty,\epsilon_{0}=0)&\approx&
-\frac{1}{\pi\,l}+\frac{l\,M}{2\pi\,\rho^2},\,
\,
E(\rho\rightarrow\infty,\epsilon_{0}=0)\approx-\frac{2\rho}{l}+\frac{l\,M}{\rho}
,\nonumber\\
M(\rho\rightarrow\infty,\epsilon_{0}=0)&\approx&2M-2\frac{\rho^2}{l^2}.
\end{eqnarray}
Although the expression of $M(\rho,\epsilon_{0}=0)$ holds in the
whole spacetime and not only
in the boundary at spatial infinity, the approximation sign $\approx$
is used instead  of the equality $=$ to be consistent
with the point under consideration.The reference energy density to be used in this work is
 the one corresponding to the anti--de Sitter metric with parameter $M_{0}$,
 $\epsilon_{0}(M_{0})=-\frac{1}{\pi\rho}\sqrt{\frac{\rho^2}{l^2}-M_{0}},\,\epsilon_{0\mid\infty}(M_{0})\approx
 -\frac{1}{\pi\,l}+\frac{l\,M_{0}}{2\pi\,\rho^2},$
 then the expansions of the physical characteristics at
 spatial infinity, $\rho \rightarrow \infty$,
 are given  as
 \begin{eqnarray}\label{enerBTZapproxEps1}
 \epsilon(\rho\rightarrow\infty, \epsilon_{0\mid\infty}(M_{0})) &&\approx\frac{l}{2\pi \,\rho^2}(M-M_{0}),
 E(\rho\rightarrow\infty, \epsilon_{0\mid\infty}(M_{0}))\approx\,l\,\frac{(M-M_{0})}{\rho},\nonumber\\
 M(\rho\rightarrow\infty, \epsilon_{0\mid\infty}(M_{0}))&&\approx\,M-M_{0}.
 \end{eqnarray}
Thus, comparing the quasi local energy and mass with the corresponding quantities associated to
the AdS solution one sees that energies vanish while mass stays finite as the radial coordinate approaches
infinity. For the proper AdS reference metric one has to equate $M_{0}=-1$.

 \end{document}